\documentclass[twocolumn,showpacs,preprintnumbers,amsmath,amssymb,superscriptaddress]{revtex4}
\usepackage{calc}
\usepackage{graphicx}% Include figure files
\usepackage{bm}% bold math
\usepackage{txfonts}
\usepackage{color}
\begin{document}

\title{Facilitated diffusion of proteins on chromatin}

%\title{Optimal narrow escape time for surface mediated intermittent diffusion}

\author{O. B\'enichou}
\affiliation{Laboratoire de Physique Th\'eorique de la Mati\`ere Condens\'ee
CNRS-UPMC, 4 Place Jussieu, 75255
Paris Cedex}

\author{C. Chevalier}
\affiliation{Laboratoire de Physique Th\'eorique de la Mati\`ere Condens\'ee
CNRS-UPMC, 4 Place Jussieu, 75255
Paris Cedex}

\author{B.Meyer}
\affiliation{Laboratoire de Physique Th\'eorique de la Mati\`ere Condens\'ee
CNRS-UPMC, 4 Place Jussieu, 75255
Paris Cedex}

\author{R. Voituriez}
\affiliation{Laboratoire de Physique Th\'eorique de la Mati\`ere Condens\'ee
CNRS-UPMC, 4 Place Jussieu, 75255
Paris Cedex}

\date{\today}

\begin{abstract}
 We present a theoretical model of facilitated diffusion of proteins in the cell nucleus. This model, which takes into account the successive  binding and unbinding events of proteins to DNA,   relies on a fractal description of the chromatin which has been recently evidenced experimentally. Facilitated diffusion is shown quantitatively to be favorable for a fast  localization of a target locus by a transcription factor, and  even to enable  the minimization of the search time by tuning the affinity of the transcription factor with DNA. This study shows the robustness of the facilitated diffusion mechanism, invoked so far only for linear conformations of DNA.
\end{abstract}

%\pacs{Valid PACS appear here}

\maketitle

\def\aver#1{{\left<{#1}\right>}}
\def\Rset{\hbox{{I\kern -0.2em R}}}
\def\rset{\hbox{{\tiny\rm I\kern -0.2em R}}}
\def\udm{{1\over2}}
\def\emt{{e^{-\alpha\t}}}
\def\emdt{{e^{-2\alpha\t}}}
\def\expo#1{{e^{^{{#1}}}}}
\def\kb{{k_{\hbox{\tiny B}}}}
\def\Te{{T_e}}
\def\kT{{\kb\Te}}
\def\capa{{\mathcal C_{\rho_H}}}
\def\capax{{\mathcal C_{x}}}
%Macros de Yann
\renewcommand{\epsilon}{\varepsilon}
\renewcommand{\leq}{\leqslant}
\renewcommand{\geq}{\geqslant}
%Pour ce calcul
\newcommand{\om}{{\boldsymbol{\omega}}}
\def\r{{\mathbf{r}}}
\def\R{\mathbf{R}}%%%%%%%%%%%%%%%%%%%%%%\input{../Brownien2D/articleB2D13.tex}
%%%%%%%%%%%%%
\renewcommand{\Rset}{\mathbb{R}}

The now well established theory of facilitated diffusion explains how DNA-binding proteins can in principle find their target sites on DNA efficiently. This model describes search trajectories as alternating phases of free diffusion in the bulk cytoplasm and 1-dimensional diffusion along the DNA strand, called sliding, which is made possible by sequence independent interactions  of proteins with DNA. Since the seminal work  \cite{Berg:1981}, such pathways have been evidenced experimentally both in vivo \cite{Elf:2007} and in vitro \cite{Bonnet:2008,Loverdo:2009a,Broek:2008} thanks to single molecule technics, and theoretical aspects have been refined \cite{Hu:2006a,Sheinman:2009gf, Benichou:2009a,PRL_dahirel_paillusson}, in particular highlighting that such  strategies can  minimize the search time for a target site by a proper   tuning of the protein/DNA interaction \cite{Slutsky:2004a,Coppey:2004,Lomholt:2005,Eliazar:2007a}.

All these theoretical  approaches rely on a schematic description of DNA as a 1-dimensional linear chain along which a protein can diffuse, surrounded  by an homogeneous medium in which the protein performs regular diffusion. More recently, crowding effects have been incorporated in these models both for the sliding  \cite{Li:2009} and the bulk cytoplasmic phases \cite{Loverdo:2009a}, leading to more realistic descriptions of gene regulation kinetics in prokaryots.  

 However, such models are  clearly inapplicable to eukaryots, in which the DNA is packed in the cell nucleus \cite{Bystricky:2004uq} and forms  a complex structure called chromatin which is far from a simple 1-dimensional  chain. Even if a full bottom-up description of the in vivo DNA organization remains out of reach, theoretical ideas \cite{crumpleddna}, and now growing experimental evidences  indicate that the chromatin has a hierarchized architecture  which displays fractal properties at least over the 100 nm -- 10 $\mu$m range. Indeed, textural image analysis, neutron scattering \cite{Lebedev:2008}, rheology technics \cite{Bancaud:2009} and more recently the Hi-C method \cite{Lieberman-Aiden:2009fk} revealed independently a fractal structure of the chromatin characterized by a fractal dimension $d_f$ which was found in the range $2.2-3$. Despite this complex structure of the chromatin, the switching dynamics of proteins between a DNA bound state and a freely diffusing state, which characterizes facilitated diffusion in prokaryots,  seems to be also at work in the nucleus, as evidenced on the examples of histons,  high-mobility group proteins, and more generally chromatin binding proteins \cite{Misteli:2001zt}. This naturally raises the questions of determining whether the classical facilitated diffusion mechanism can be  efficient also in the complex nuclear environment, and whether it can be used to regulate and optimize gene expression in eukaryots. This letter presents a first theoretical model which quantitatively addresses these two questions.

\begin{figure}[h!]
   \centering
   \includegraphics[scale=0.13]{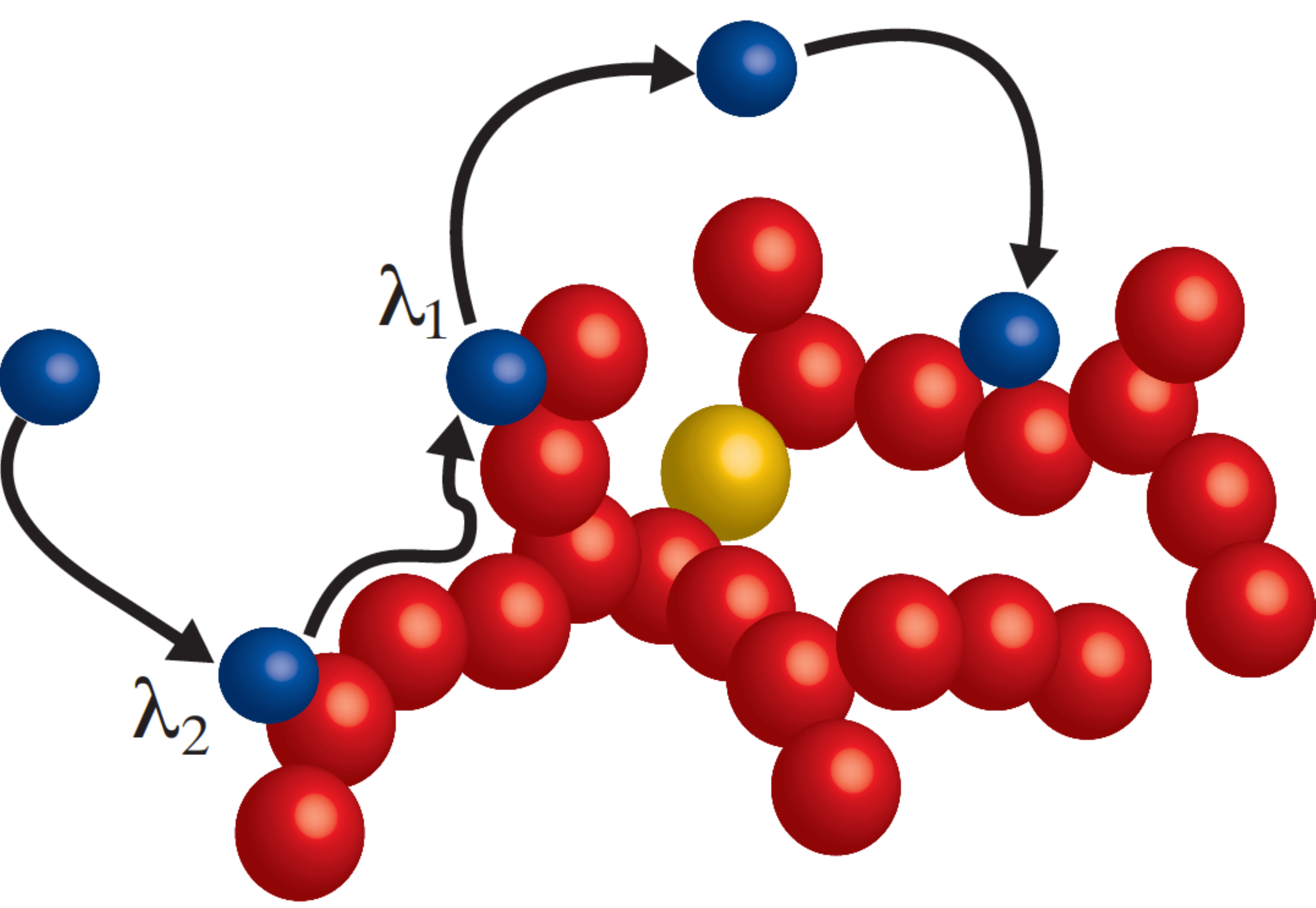}
   \caption{\small{Facilitated diffusion on chromatin: A chromatin binding protein (blue) searches for a target locus (yellow) on chromatin (red). The binding (resp.  unbinding) rate is denoted by  $\lambda_2$  (resp. $\lambda_1$).}}
   \label{picture}
\end{figure}

At the theoretical level, modelling facilitated diffusion in the cell nucleus raises two problems: (i) first, to take into account the switching dynamics of the protein between a state bound to the chromatin and a freely diffusing state in the nucleoplasm, and (ii) second to model the diffusion phase of a protein bound to a complex structure such as  chromatin. Point (i) has been studied in the context of intermittent search strategies \cite{Obenichou:2008}, and general methods to calculate mean search times for intermittent trajectories have been developed. On the other hand, the full distribution of the first-passage time (FPT) for diffusion in fractals has recently been obtained in \cite{BenichouO.:2010} and  enables to tackle point (ii) under the assumption, backed by experiments \cite{Lebedev:2008,Bancaud:2009,Lieberman-Aiden:2009fk}, that the chromatin is fractal. 

In this letter, we gather and extend these new tools to develop  a theoretical  model of facilitated diffusion in the cell nucleus. More precisely, we calculate analytically the mean search time for a target for a protein which alternates diffusion phases on the chromatin which is assumed to have a fractal structure, and free diffusion phases in the nucleoplasm (see Fig. \ref{picture}). Under these hypothesis, we show quantitatively  that facilitated diffusion in eukaryots  can significantly speed up the search process, and that it enables to minimize the search time by tuning the affinity of the protein with DNA. These results are qualitatively similar to the case of prokaryots, and suggest that facilitated diffusion is a robust mechanism. At the theoretical level,  this study yields as a by-product the calculation of the distribution of the FPT averaged over the starting point for a particle diffusing in a fractal structure, which remained a challenge in the field \cite{Kozak:2002,Haynes:2008,Agliari:2008,agliari-2009}.

{\it Search time distribution for simple diffusion in a fractal medium}. We first consider a protein which remains in an adsorbed state and diffuses on the chromatin, which is modelled by a  
 discretized  domain $\mathcal D$ of volume $N$, and, following experimental observations \cite{Lebedev:2008,Bancaud:2009,Lieberman-Aiden:2009fk}, is characterized by a fractal dimension $d_f$ 
and a typical size $\displaystyle{R \propto N^{1/d_f}}$.  We calculate here  the distribution of the search time, defined as the FPT at the target, averaged over the starting position of the protein. This first technical step is necessary to address the problem of facilitated diffusion discussed in the next paragraph; besides, it is an important  theoretical question.   The protein of position $\mbox{\bf{r}}(t)$,  is assumed to perform  a symmetric  nearest neighbor
random walk on the chromatin with a constant hopping rate (set to 1). To  account for the complex organization of  DNA-DNA contact points the chromatin cannot be described as a linear chain: it is  effectively branched even if the DNA is linear, yielding a connectivity potentially larger than 2 which takes into account intersegmental transfer. The resulting 
dynamics is characterized by the walk dimension $d_w$ defined 
through the scaling of the mean square displacement with time: 
$\displaystyle{\aver{\mbox{\bf{r}}^2(t)}\propto t^{2/d_w}}$, and the nuclear membrane which bounds the chromatin is assumed to act as reflecting walls. As we proceed to show, the search time distribution is independent of these microscopic details of the chromatin conformation, and is governed only by its larger scale properties which are characterized by $d_f$ and $d_w$.

We denote by $W_{ji}(t)$ the propagator, \textit{i.e.} 
the probability that the protein, starting at site $i$ at $\displaystyle{t=0}$, 
is at site $j$ at time $t$, and  write $W_{j}^{\mbox{\tiny{stat}}}$ for the stationary probability at site $j$. 
We will make use of the pseudo-Green function of the walk defined by
$H_{ji}=\int_{t=0}^{\infty}(W_{ji}(t)-W_j^{\mbox{\tiny{stat}}})dt$, and the Laplace transform of a generic function $f(t)$ will be denoted by $\hat f(s)$.
We are here interested in the  global FPT (GFPT) at a given target site $T$, which  
is the FPT at site $T$ {\it averaged} over the starting  site $S$ with weight  $W_{S}^{\mbox{\tiny{stat}}}$. 
We thus  define the probability density $\Phi_T$ of the GFPT by 
$\Phi_T(t)= \sum_{j=1}^{N}W^{\mbox{\tiny{stat}}}_{j}\, P_{Tj}(t)$,  
where  $P_{Tj}$ is the probability density of the FPT at $T$ starting from {\it a given}  site  $j$.

General expressions have been derived for the 
first moment of both the FPT \cite{package_nature} and the   
 GFPT \cite{Tejedor:2009} (see also \cite{Kozak:2002,Haynes:2008,Agliari:2008,agliari-2009} for specific examples), and more recently, the higher FPT moments have been determined in the  large-volume limit $N \gg 1$ \cite{BenichouO.:2010}. In the case of non-compact exploration $(d_w<d_f)$, it reads
\begin{equation}
\label{eq:FPTMom}
\aver{\tau^n_{TS}}= n! \, \aver{\tau_T}^n 
\frac{H_{TT}-H_{TS}}{H_{TT}} ,
\end{equation}
where  $\aver{\tau_T}=H_{TT}/W_T^{\mbox{\tiny{stat}}}$ is the mean GFPT \cite{Tejedor:2009} .
Using next that $\displaystyle{H_{ji}\, W_{i}^{\mbox{\tiny{stat}}}= H_{ij} \, W_{j}^{\mbox{\tiny{stat}}}}$, deduced from detailed balance, and averaging over $S$, we obtain 
$\aver{\tau^n_T}= n! \, \aver{\tau_T}^n$, 
from which it can be deduced immediately that  the GFPT distribution is a simple exponential of mean  $\aver{\tau_T}$.

%\begin{equation}
%\label{eq:GFPTPDFncmarg}
%\Phi_T(t)=\frac{1}{\aver{\tau_T}}\, \exp \left(-\frac{t}{\aver{\tau_T}}\right). 
%\end{equation}
In the compact case $(d_w>d_f)$ however, it can be shown that due to a stronger  dependence of  $\aver{\tau^n_{TS}}$ on  $S$ \cite{package_nature,BenichouO.:2010},  the average over $S$ must be taken before the large $N$ limit, which makes the asymptotic form of \cite{BenichouO.:2010} unusable for this purpose. Alternatively, one can make use   of the renewal equation \cite{Redner:2001}
 which reads in Laplace space $\hat P_{TS}(s)=\hat W_{TS}(s)/\hat W_{TT}(s)$. Using the symmetry relation $\displaystyle{W_{ji}(t)\, W_{i}^{\mbox{\tiny{stat}}}= W_{ij}(t) \, W_{j}^{\mbox{\tiny{stat}}}}$, the average over $S$ can be taken and  yields an  exact expression of the GFPT distribution : $
\hat \Phi_T(s)=W_{T}^{\mbox{\tiny{stat}}}/(s\hat W_{TT}(s)).$
Taking next the large volume limit, the propagator $\hat W_{TT}(s)$ can be evaluated using the O'Shaughnessy and Procaccia formalism \cite{OShaughnessy:1985}, which leads   to
\begin{equation}
\label{eq:LapTransfGFPTPDFnoninterm}
%\begin{split}
\hat \Phi_T(s)= \, \left(\frac{4}{A\,\aver{\tau_T} \,s}\right)^{\nu}\, 
\frac{\Gamma(1+\nu)}{\Gamma(1-\nu)}\, 
%\\&
\times \frac{I_{\nu}\left(\sqrt{A \,\aver{\tau_T}\,s}\,\right)}
{I_{-\nu}\left(\sqrt{A \,\aver{\tau_T}\,s}\,\right)}, 
%\end{split}
\end{equation}
where $\displaystyle{\nu=d_f/d_w}$, 
%\begin{equation}
% \label{eq:defA}
$A=2 (1-\nu ^2)/\nu$, and $I_\nu$ (and later $J_\nu$) denote Bessel functions.  % (s+\lambda_1),  
%\end{equation}
Introducing  the rescaled variable  $\displaystyle{\theta=t/\aver{\tau_T}}$, one finally obtains by inverse-Laplace transforming (\ref{eq:LapTransfGFPTPDFnoninterm})  : 
\begin{eqnarray}
\label{eq:GFPTPDFthetaSummary}
Q_T(\theta) &=& \exp(-\, \theta)   
\ \ \ \ \ \ \ \ \ \ \ \ \ \ \ \ \ \ \ \ \ \ \ \ \ \ \ \ \ \ \ \ \ \ \ \ \  (d_w<d_f) \nonumber \\
&=& \frac{2^{2\nu}\, \nu}{(1-\nu^2)} 
\, \frac{\Gamma(1+\nu)}{\Gamma(1-\nu)}\, 
\sum_{k=0}^{\infty}\, \alpha_k^{1-2\nu}\, \frac{J_{\nu}(\alpha_k)}{J_{1-\nu}(\alpha_k)}\nonumber \\
& &\ \ \times \,\exp \left(-\, \frac{\alpha_k^2\, \nu}{2 \,(1-\nu^2)} \, \theta \right)   
\,  \ \ \ \ \  \ \ \ \ \ (d_w>d_f),\nonumber \\%\nonumber \\ 
%&=&� \exp(-\, \theta) 
%\, \ \ \ \ \ \ \ \ \ \ \ \ \ \ \ \ \ \ \ \ \ \ \ \ \ \ \ \ \ \ \ \ \ \ \ \ (d_w=d_f).  \nonumber \\ 
\end{eqnarray}
where the $\alpha_k$'s are the real zeros of $J_{-\nu}$. 
\begin{figure}[h!]
   \centering
   \includegraphics[scale=0.33]{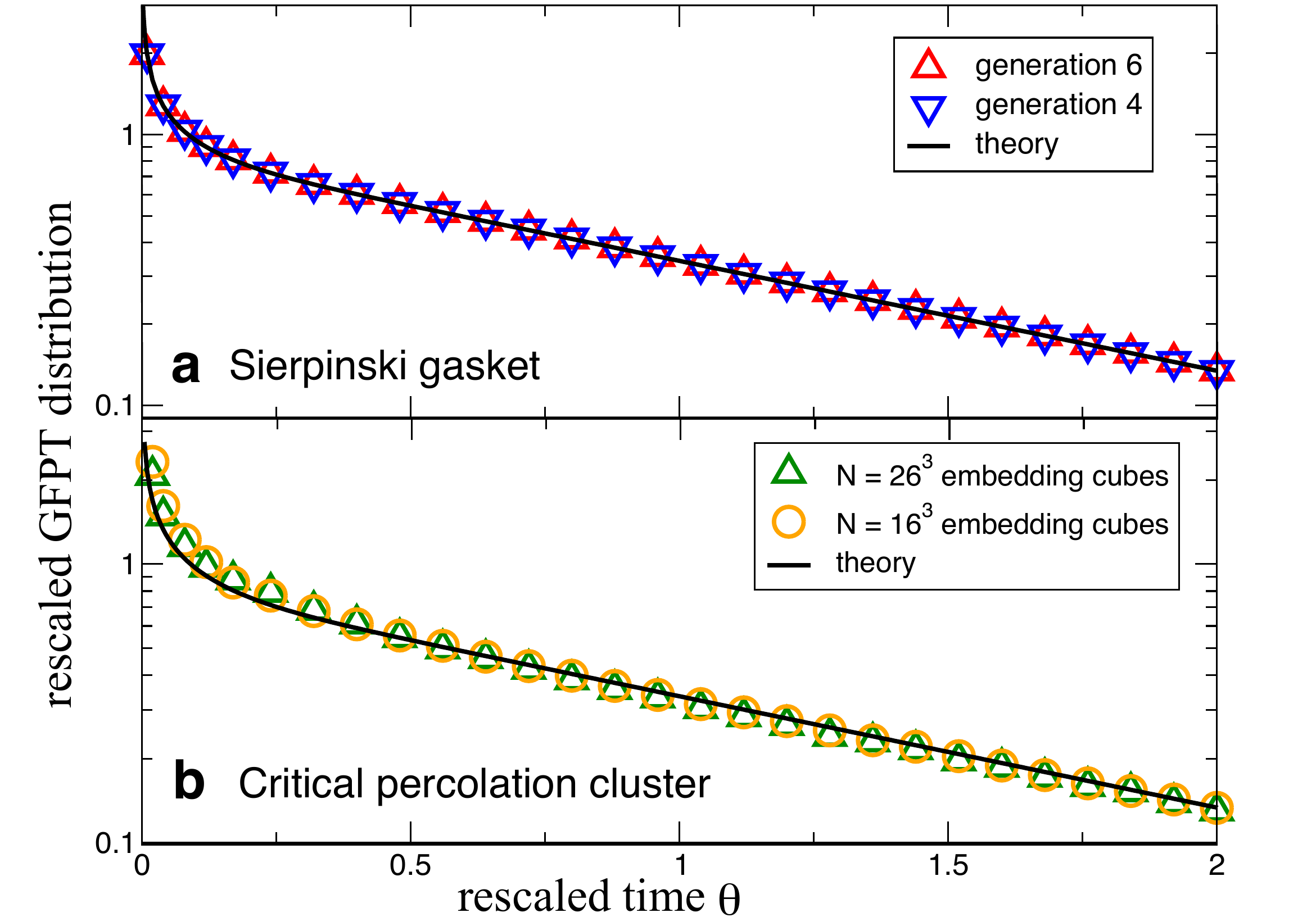}
   \caption{\small{Distribution of the GFPT for diffusion in fractal medium in the compact case. Numerical simulations for different system sizes $N$ (symbols) are plotted against the theoretical prediction of Eq. \ref{eq:GFPTPDFthetaSummary} (plain lines). ${\bold a}$ Sierpinsky gasket (target at the apex) and ${\bold b}$ critical bond percolation cluster embedded in a 3D cubic lattice (average taken over random targets and cluster realizations).}}
   \label{fig:GFPTSierp}
  
\end{figure}

%in agreement with exact results obtained for the second moment of the GFPT \cite{Haynes:2008}

This very general result, confirmed by numerical simulations on various fractal sets (see fig. \ref{fig:GFPTSierp} and \cite{si}) shows that the GFPT distribution  takes a universal form  indexed by $d_f$ and $d_w$ only for any fractal structure, independently of its microscopic details. In particular, it shows the applicability of our approach to chromatin under the assumption that it is fractal at least at sufficiently large scale. As illustrative examples, simulations were performed on structures such as critical percolation clusters, which capture the large scale properties of nuclear DNA organization :  they are (i)  fractals characterized by well defined $d_w$ and $d_f$, (ii) naturally embedded in euclidean space and (iii) disordered.  These fractals can therefore be seen as minimal models of chromatin beyond the linear chain description.

 %It shows that the GFPT distribution is fully determined as soon as its first moment $\aver{\tau_T}$ is known (which is the case for many examples \cite{Kozak:2002,Haynes:2008,Agliari:2008,agliari-2009}) , 
%and puts forward the differences between compact and non compact transport. First, the relevant time scale  behaves in the large system size limit  as   $\aver{\tau_T}\propto N$ in the non compact case and as $\aver{\tau_T}\propto N^{d_w/d_f}$ in the compact case \cite{Tejedor:2009}, reminiscent of   a much slower exploration. Second,
%the distribution is a single exponential only in the non compact case, while an infinite set of time scales is expected in the compact case. This shows that  the widely used quasi-chemical approximation of a single exponential decay \cite{Grigoriev:2002a}  is valid only in the case of non-compact exploration, and only after averaging over the initial condition of the reactant. 

{\it Search time distribution  for facilitated diffusion in a fractal medium}. Following the classical picture of facilitated diffusion, we now consider that  the protein can desorb from the chromatin with rate $\lambda_1$, and then  freely diffuse in the nucleoplasm before rebinding to the chromatin (see Fig. \ref{picture}).  In this first approach, we adopt a mean field treatment of the phases of free diffusion : we assume that the duration of such a phase is exponentially distributed with mean $\tau_2=1/\lambda_2$, and further suppose that the protein rebinds at a position which is uniformly distributed on the chromatin. 
We  determine in this paragraph the mean time necessary for the protein starting 
at a random position on the chromatin to reach a target locus on the chromatin for  the first time. 
We denote by $\aver{\mbox{\bf{T}}_T}$ this mean GFPT for facilitated diffusion,  and by $F_T(t)$ the GFPT probability density. 
 Using tools developed in the context of intermittent search strategies \cite{Coppey:2004,Obenichou:2008} , it can be shown that  the Laplace transform  $\hat F_T$ is expressed in terms of the distribution $\hat \Phi_T$ of the GFPT for simple diffusion on the chromatin, which is given by  Eq.(\ref{eq:LapTransfGFPTPDFnoninterm})  : 
\begin{equation}
 \label{eq:genexprhatF}
\hat F_T(s)=\hat \Phi_T(\lambda_1+s) 
\left[1-\frac{1-\hat \Phi_T(\lambda_1+s)}{\left(1+\frac{s}{\lambda_1} \right)\left( 1+\frac{s}{\lambda_2}\right)} \right]^{-1}.
\end{equation}
Assuming that diffusion on the chromatin is compact (as in most examples of fractals embedded in $3D$ space \cite{D.Ben-Avraham:2000}) an using expression (\ref{eq:LapTransfGFPTPDFnoninterm}) for $\hat \Phi_T$ finally yields an explicit expression of the Laplace transformed distribution of the GFPT: 
\begin{equation}
 \label{eq:hatFcompact}
\begin{split}
&\hat F_T(s)=(s+\lambda_1) (s+\lambda_2) I_{\nu }(x_s)
\\& 
\times \left[
\frac{\Gamma (1-\nu )}{4^\nu\Gamma (1+\nu)}x_s^{2\nu} s(s+\lambda_1+\lambda_2) 
I_{-\nu}(x_s)+\lambda_1 \lambda_2 I_{\nu}(x_s)\right]^{-1}
\end{split}
\end{equation}
where $x_s\equiv\sqrt{A \aver{\tau_T}(s+\lambda_1)}$. One can check that the classical result of facilitated diffusion on a 1-dimensional DNA \cite{Coppey:2004,Obenichou:2008} is recovered for $d_f=1$ and $d_w=2$.
The mean GFPT $\aver{\mbox{\bf{T}}_T}$ is then readily obtained  by writing  
\begin{equation}
 \label{eq:defGMFPTinterm}
\aver{\mbox{\bf{T}}_T}=-\left(\frac{\partial \hat F_T}{\partial s}\right)_{s=0}=\left(\frac{1}{\lambda_1}+\frac{1}{\lambda_2} \right) 
\frac{1-\hat \Phi_T(\lambda_1)}{\hat \Phi_T(\lambda_1)},
\end{equation}
which finally yields the central result of this paper: 
\begin{equation}
 \label{eq:GMFPTintermcompact}
\aver{\mbox{\bf{T}}_T}=\left(\frac{1}{\lambda_1}+\frac{1}{\lambda_2} \right)
\left[\frac{\Gamma (1-\nu )}{4^\nu\Gamma (1+\nu)} 
\frac{x_0^{2\nu}I_{-\nu }(x_0)}{I_{\nu }(x_0)}-1\right], 
\end{equation}
with $x_0\equiv x_{s=0}$.
\begin{figure}[h!]
   \centering
   \includegraphics[scale=0.3]{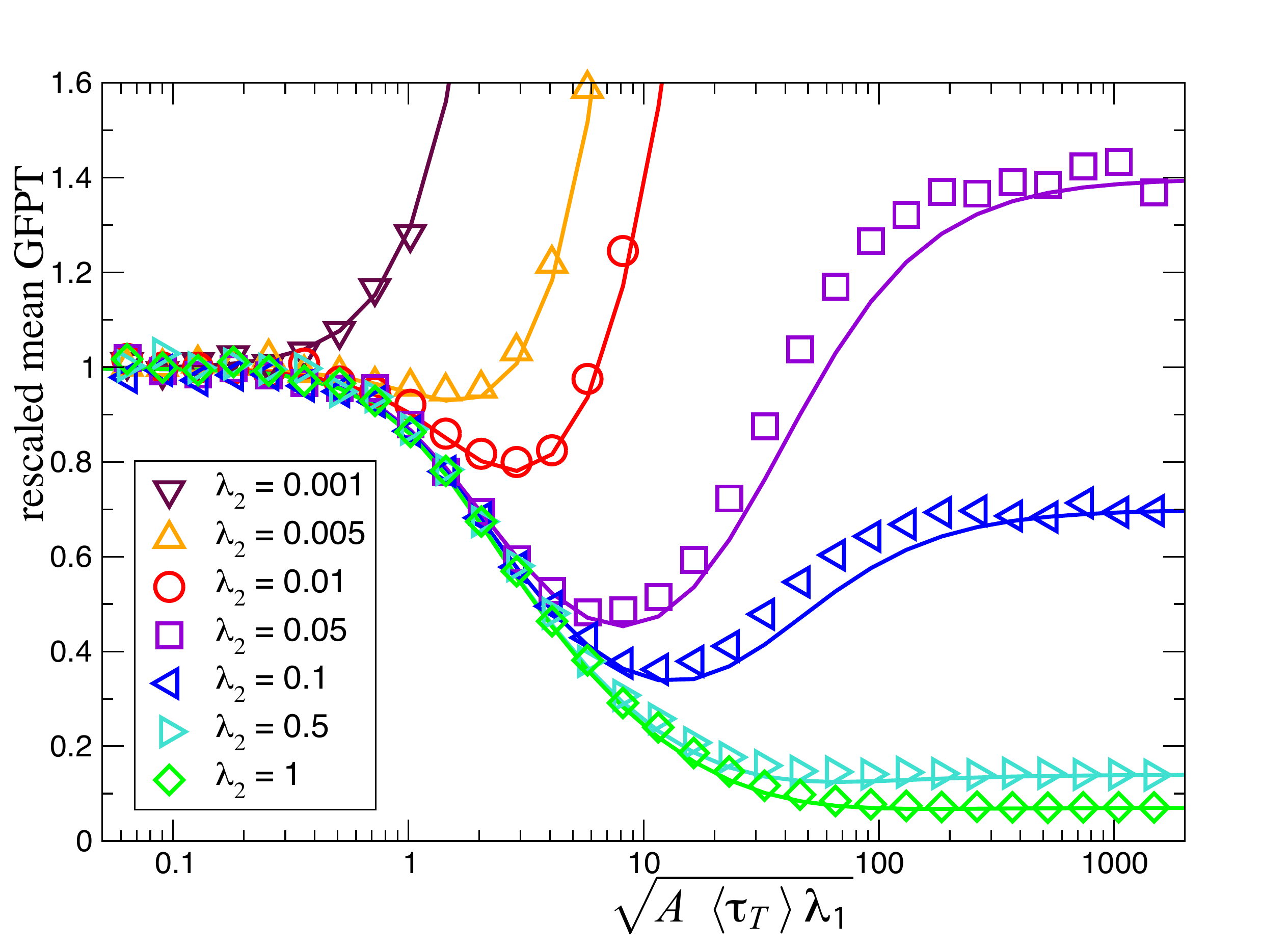}
	%\vspace{-0.5cm}
   \caption{\small{Mean search time for facilitated diffusion  on a Sierpinski gasket of generation 5 ($N=366$) with a target at the apex. Numerical simulations (symbols) and theoretical prediction (Eq. \ref{eq:GMFPTintermcompact}, plain lines) of the mean GFPT , rescaled by its value for $\lambda_1=0$, is plotted as a function of  $x_0=\sqrt{A \,\aver{\tau_T}\lambda_1}$ for various $\lambda_2$.}}
   \label{fig:GMFPTintermSierp}
\end{figure}
\begin{figure}[h!]
   \centering
   \includegraphics[scale=0.3]{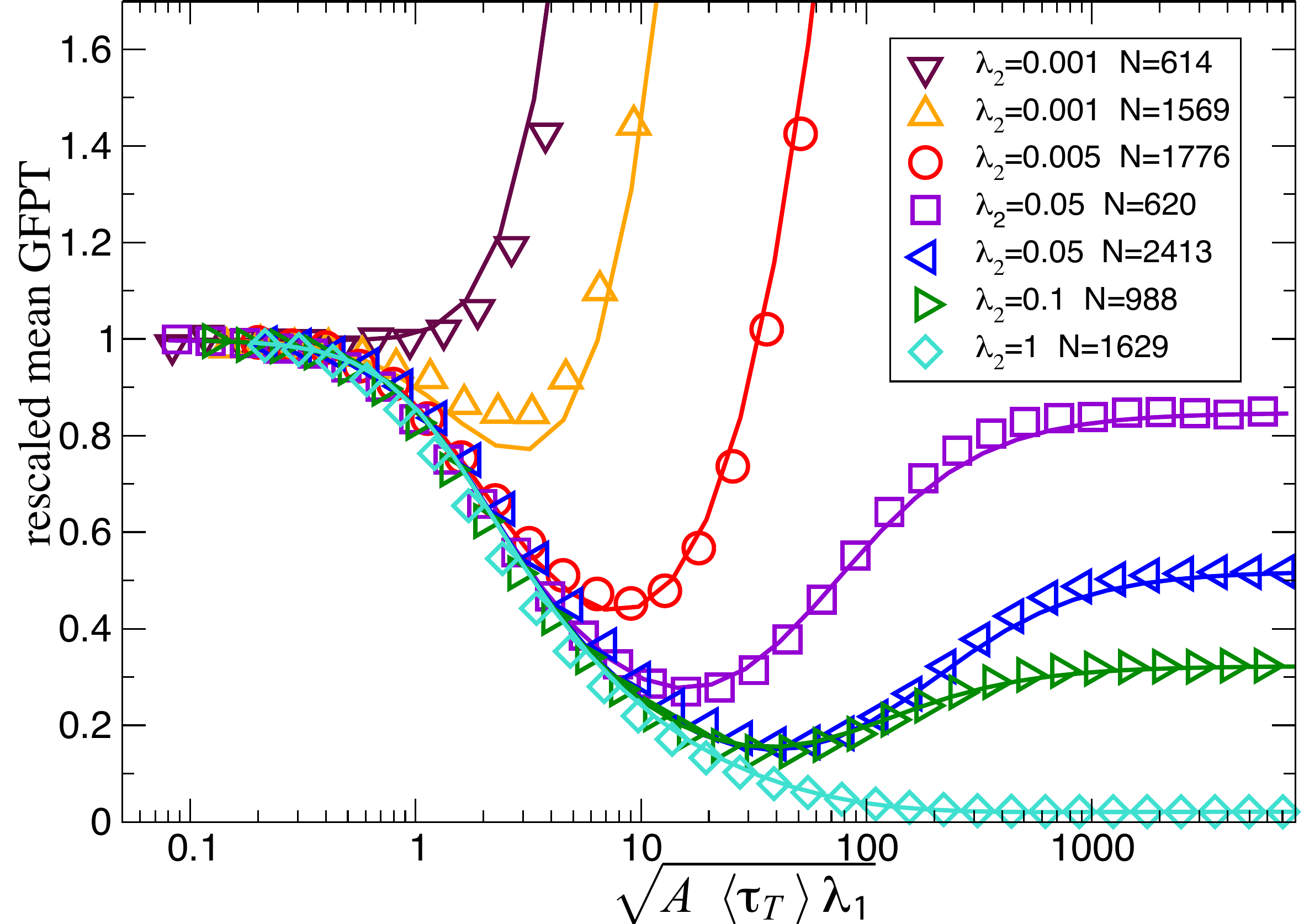}
	%\vspace{-0.5cm}
   \caption{\small{Mean search time for facilitated diffusion  on critical bond percolation clusters embedded in a 3D cubic lattice with a randomly located target. Numerical simulations (symbols) and theoretical prediction (Eq. \ref{eq:GMFPTintermcompact}, plain lines) of the mean GFPT , rescaled by its value for $\lambda_1=0$, is plotted as a function of  $x_0=\sqrt{A \,\aver{\tau_T}\lambda_1}$ for various $\lambda_2$ and system size $N$.}}
   \label{fig:GMFPTintermperco}
\end{figure}
This exact expression of the mean search time for facilitated diffusion in a fractal medium, validated by numerical simulations on various examples of fractals   which mimic the large scale properties of chromatin (see Figs \ref{fig:GMFPTintermSierp}, \ref{fig:GMFPTintermperco} and \cite{si})  can be simplified as follows in the large   $N$ limit   : 
\begin{equation}
 \label{eq:GMFPTlargevolume}
\aver{\mbox{\bf{T}}_T}\sim \aver{\mbox{\bf{T}}_T}_{\infty} \equiv \left(\frac{1}{\lambda_1}+\frac{1}{\lambda_2} \right)
\frac{x_0^{2\nu}\Gamma (1-\nu )}{4^\nu\Gamma (1+\nu)}.  
%\frac{I_{-\nu }\left(\sqrt{A \,\lambda_1}\right)}{I_{\nu }\left(\sqrt{A \,\lambda_1}\right)}. 
\end{equation}
Before commenting on this result, note that a limit distribution $G_T$  of the GFPT can be obtained by  considering the rescaled time $\displaystyle{\Theta=t/ \aver{\mbox{\bf{T}}_T}_{\infty}}$. Denoting $\displaystyle{U=s \aver{\mbox{\bf{T}}_T}_{\infty}}$ the Laplace variable associated to $\Theta$, and  
writing $\displaystyle{\hat G_T(U)=\hat F_T(U/\aver{\mbox{\bf{T}}_T}_{\infty})}$, 
we find using (\ref{eq:hatFcompact}) in the large  volume limit $\hat G_T(U) \sim (1+U)^{-1}$, which yields immediately the simple  exponential form $G_T(\Theta) \sim \exp (-\Theta).$

% \begin{figure}[h!]
  % \centering
   %\includegraphics[scale=0.3]{fig5}
	%\vspace{-0.5cm}
%   \caption{\small{GFPT PDF  for intermittent transport on a Sierpinsky gasket}}	
   %\label{fig:GFPTintermasympt}
%\end{figure}

{\it Optimal search time.} We now comment on previous results and  focus on the minimization of the mean search time  (\ref{eq:GMFPTintermcompact}) for facilitated diffusion. First note that $\aver{\mbox{\bf{T}}_T}_{\infty}\propto N$ in the large volume limit, which shows immediately that facilitated diffusion if faster than diffusion alone, which would yield a search time scaling as $N^{d_w/d_f}$  \cite{Tejedor:2009}.
Actually, the search time can be minimized as a function of the desorption rate $\lambda_1$, as soon as the adsorption rate $\lambda_2$ is large enough. 
Quantitatively, 
the function $\aver{\mbox{\bf{T}}_T}$ exhibits a minimum value for some 
$\displaystyle{\lambda_1=\lambda_1^{\mbox{\tiny{min}}}}$ 
if the value of the derivative of  $\aver{\mbox{\bf{T}}_T}$ with respect to $\lambda_1$ at $\lambda_1=0$ is 
negative. 
This sets the following condition on $\lambda_2$ : 
\begin{equation}
\label{eq:condlambda2cont}
\lambda_2 \geq \lambda_2^{\mbox{\tiny{min}}}=\frac{4 \nu(4-\nu^2)}{\aver{\tau_T} (5+2\nu) (1-\nu)^2}, 
\end{equation}
which is in practice satisfied for a large enough chromatin volume.
 %In the case of 1D normal diffusion $\displaystyle{\lambda_2^{\mbox{\tiny{min}}}=5/\aver{\tau_T}=15 D/R^2}$. 
Under this condition, a direct calculation shows that the optimal value  $\lambda_1^{\mbox{\tiny{min}}}$ can be expanded  in the large volume limit as:  
\begin{equation}
 \label{eq:lambda1minexpansion}
%\begin{split}
\frac{\lambda_1^{\mbox{\tiny{min}}}}{\lambda_2}\simeq\frac{1-\nu}{\nu}
-\frac{\Gamma^2(\nu) \sin(\pi \nu)}{\pi}\left(\frac{2 \nu^2}{\aver{\tau_T}\lambda_2 (1+\nu) (1-\nu)^2}\right)^{\nu}.
%\end{split}
\end{equation}
One recovers in particular for $d_f=1$ and $d_w=2$ the celebrated result $\lambda_1^{\mbox{\tiny{min}}}\simeq\lambda_2$ \cite{Slutsky:2004a,Coppey:2004}.

Finally, these results show that facilitated diffusion is a robust mechanism which can speed up the search for a target site even in the case of eukaryots, under the assumption that the chromatin has a fractal organization, which seems verified experimentally. We add that this approach is independent of the microscopic structure of the DNA and could also be valid to some extent  in the case of prokaryots, where the DNA, even if less densely packed than in eukaryots seems to have a rather compact organization significantly departing from a linear chain.  Using typical experimental values $d_w\simeq 3$ and $d_f\simeq 2.5$, 
one finds that the search time is minimized for $\lambda_1^{\mbox{\tiny{min}}}/\lambda_2\simeq1/5$. This suggests that the  adsorption time should be significantly larger  than the time of free diffusion to minimize the search, in contrast with the classical prediction \cite{Slutsky:2004a,Coppey:2004}. This result is qualitatively compatible with experimental findings \cite{Elf:2007} on prokaryots, even if the fractal properties of the DNA structure needs to be determined in this case.

%To conclude, we have presented a theoretical model of facilitated diffusion of proteins in the cell nucleus. This model, potentially relevant to other search problems on fractal structures,  assumes a fractal description of the chromatin which has been recently evidenced experimentally. Facilitated diffusion is shown quantitatively to be favorable for a fast  localization of a target site by a transcription factor, and can even lead to a minimization of the search time by tuning the affinity of the protein  with DNA. These findings are qualitatively close to the classical  results obtained for a linear DNA conformation, and suggest that facilitated diffusion is a robust mechanism which can be used to regulate gene expression.

%\bibliography{liste}

\end{document}